\pdfoutput=1
\documentclass[pdflatex,sn-mathphys-num]{sn-jnl}

\usepackage{graphicx}%
\usepackage{multirow}%
\usepackage{amsmath,amssymb,amsfonts}%
\usepackage{amsthm}%
\usepackage{mathrsfs}%
\usepackage[title]{appendix}%
\usepackage{xcolor}%
\usepackage{textcomp}%
\usepackage{manyfoot}%
\usepackage{booktabs}%
\usepackage{algorithm}%
\usepackage{algorithmicx}%
\usepackage{algpseudocode}%
\usepackage{listings}%

\theoremstyle{thmstyleone}%

\theoremstyle{thmstyletwo}%

\theoremstyle{thmstylethree}%

\begin{document}

\title[Achieving a 0.49\,keV threshold in the COSINE-100 dark matter experiment]{Achieving a 0.49\,keV threshold in the COSINE-100 dark matter experiment}

\author*[1,4]{\fnm{J.} \sur{Kim}}\email{rmsidekwhdk@gmail.com}
\author[2]{\fnm{N.} \sur{Carlin}}
\author[2]{\fnm{D. F. F. S.} \sur{Cavalcante}}
\author[3]{\fnm{J. Y.} \sur{Cho}}
\author[4]{\fnm{S. J.} \sur{Cho}}
\author[5]{\fnm{S.} \sur{Choi}}
\author[6]{\fnm{A. C.} \sur{Ezeribe}}
\author[2]{\fnm{L. E.} \sur{Fran{\c c}a}}
\author[4]{\fnm{O.} \sur{Gileva}}
\author*[1]{\fnm{C.} \sur{Ha}}\email{chha@cau.ac.kr}
\author[8,9,10]{\fnm{I. S.} \sur{Hahn}}
\author[4,10]{\fnm{E. J.} \sur{Jeon}}
\author[5]{\fnm{H. W.} \sur{Joo}}
\author[4]{\fnm{W. G.} \sur{Kang}}
\author[11]{\fnm{M.} \sur{Kauer}}
\author[4]{\fnm{B. H.} \sur{Kim}}
\author[4]{\fnm{D. Y.} \sur{Kim}}
\author[3]{\fnm{H. J.} \sur{Kim}}
\author[4]{\fnm{K. W.} \sur{Kim}}
\author[4]{\fnm{S. H.} \sur{Kim}}
\author[5]{\fnm{S. K.} \sur{Kim}}
\author[10,4]{\fnm{W. K.} \sur{Kim}}
\author[4,10]{\fnm{Y. D.} \sur{Kim}}
\author[4,10]{\fnm{Y. H.} \sur{Kim}}
\author[12]{\fnm{B. R.} \sur{Ko}}
\author[13]{\fnm{Y. J.} \sur{Ko}}
\author[1]{\fnm{B. C.} \sur{Koh}}
\author[3]{\fnm{D. H.} \sur{Lee}}
\author[4]{\fnm{E. K.} \sur{Lee}}
\author[10,4]{\fnm{H.} \sur{Lee}}
\author[4,10]{\fnm{H. S.} \sur{Lee}}
\author[8]{\fnm{H. Y.} \sur{Lee}}
\author[4]{\fnm{I. S.} \sur{Lee}}
\author[4]{\fnm{J.} \sur{Lee}}
\author[3]{\fnm{J. Y.} \sur{Lee}}
\author[4,10]{\fnm{M. H.} \sur{Lee}}
\author[10,4]{\fnm{S. H.} \sur{Lee}}
\author[13]{\fnm{S. H.} \sur{Lee}}
\author[1]{\fnm{Y. J.} \sur{Lee}}
\author[4]{\fnm{D. S.} \sur{Leonard}}
\author[3]{\fnm{N. T.} \sur{Luan}}
\author[2]{\fnm{V. H. A.} \sur{Machado}}
\author[2]{\fnm{B. B.} \sur{Manzato}}
\author[14]{\fnm{R. H.} \sur{Maruyama}}
\author[7]{\fnm{R. F.} \sur{Muhdi}}
\author[4]{\fnm{S. L.} \sur{Olsen}}
\author[12]{\fnm{H. K.} \sur{Park}}
\author[15]{\fnm{H. S.} \sur{Park}}
\author[16]{\fnm{J. C.} \sur{Park}}
\author[3]{\fnm{J. S.} \sur{Park}}
\author[4]{\fnm{K. S.} \sur{Park}}
\author[4]{\fnm{K.} \sur{Park}}
\author[3]{\fnm{S. D.} \sur{Park}}
\author[2]{\fnm{R. L. C.} \sur{Pitta}}
\author[7]{\fnm{H.} \sur{Prihtiadi}}
\author[17]{\fnm{C.} \sur{Rott}}
\author[4]{\fnm{K. A.} \sur{Shin}}
\author[16]{\fnm{M. K.} \sur{Son}}
\author[3]{\fnm{L. T.} \sur{Truc}}
\author[18]{\fnm{L.} \sur{Yang}}
\author[4]{\fnm{G. H.} \sur{Yu}}

\affil[1]{\orgname{Chung-Ang University}, \orgaddress{\city{Seoul}, \country{Republic of Korea}}}
\affil[2]{\orgname{University of S\~{a}o Paulo}, \orgaddress{\city{Sao Paulo}, \country{Brazil}}}
\affil[3]{\orgname{Kyungpook National University}, \orgaddress{\city{Daegu}, \country{Republic of Korea}}}
\affil[4]{\orgname{Center for Underground Physics, Institute for Basic Science}, \orgaddress{\city{Daejeon}, \country{Republic of Korea}}}
\affil[5]{\orgname{Seoul National University}, \orgaddress{\city{Seoul}, \country{Republic of Korea}}}
\affil[6]{\orgname{University of Sheffield}, \orgaddress{\city{Sheffield}, \country{United Kingdom}}}
\affil[7]{\orgname{Universitas Negeri Malang}, \orgaddress{\city{Malang}, \country{Indonesia}}}
\affil[8]{\orgname{Center for Exotic Nuclear Studies, Institute for Basic Science}, \orgaddress{\city{Daejeon}, \country{Republic of Korea}}}
\affil[9]{\orgname{Ewha Womans University}, \orgaddress{\city{Seoul}, \country{Republic of Korea}}}
\affil[10]{\orgname{University of Science and Technology}, \orgaddress{\city{Daejeon}, \country{Republic of Korea}}}
\affil[11]{\orgname{University of Wisconsin-Madison}, \orgaddress{\city{Madison}, \country{USA}}}
\affil[12]{\orgname{Korea University}, \orgaddress{\city{Sejong}, \country{Republic of Korea}}}
\affil[13]{\orgname{Jeju National University}, \orgaddress{\city{Jeju}, \country{Republic of Korea}}}
\affil[14]{\orgname{Yale University}, \orgaddress{\city{New Haven}, \country{USA}}}
\affil[15]{\orgname{Korea Research Institute of Standards and Science (KRISS)}, \orgaddress{\city{Daejeon}, \country{Republic of Korea}}}
\affil[16]{\orgname{Chungnam National University}, \orgaddress{\city{Daejeon}, \country{Republic of Korea}}}
\affil[17]{\orgname{University of Utah}, \orgaddress{\city{Salt Lake City}, \country{USA}}}
\affil[18]{\orgname{University of California San Diego}, \orgaddress{\city{La Jolla}, \country{USA}}}

\abstract{
The COSINE-100 experiment searched for weakly interacting massive particles~(WIMPs) with NaI(Tl) crystals at the Yangyang Underground Laboratory. Using its full 6-year dataset, we lower the analysis threshold to 0.49\,keV. 
A multi-layer perceptron classifier separates scintillation events from photomultiplier tube noise. It is trained with approximately $3\times10^{6}$ simulated scintillation waveforms per crystal, separately for groups of events with similar numbers of photoelectron clusters. 
The selection keeps more than 10\,\% of scintillation events at 0.49\,keV while limiting noise contamination to below 1\,\%. 
The selection efficiency is validated with $^{22}$Na-calibration data and internal $^{22}$Na events, and the selection performance is stable over the six years of data taking. 
With this threshold, the median expected upper limit on the spin-dependent WIMP--proton cross section reaches $1.1 \times 10^{-36}\,\text{cm}^{2}$ at $\rm m_{WIMP}=2.5\,GeV/c^{2}$, probing previously unexplored parameter space in the 2.0--2.8\,GeV/$c^{2}$ mass range.
}

\maketitle

\section{Introduction}\label{sec1}

Dark matter~(DM) is estimated to constitute approximately 26.5\,\% of the total mass--energy content of the Universe, and a wide range of astrophysical and cosmological observations over the past several decades provide strong evidence for its existence~\cite{Planck:2018nkj, Bertone:2016nfn}. Numerous theoretical models have been proposed to describe DM particles and their interactions with standard model particles. Among the well-motivated candidates, WIMPs have been extensively studied, as they can naturally account for the observed relic abundance through thermal freeze-out in the early Universe~\cite{Lee:1977ua, Jungman:1995df}. Despite extensive experimental efforts, no unambiguous DM signal has yet been observed~\cite{Misiaszek:2023sxe, ParticleDataGroup:2024cfk}, with the notable exception of the DAMA/LIBRA experiment.

The DAMA/LIBRA collaboration has reported an annual modulation signal in NaI(Tl) detectors and attributes it to DM interactions~\cite{Bernabei:1998fta, Bernabei:2013xsa, Bernabei:2018jrt, Bernabei:2021kdo}. Under standard assumptions about the DM halo and interactions, the corresponding parameter space has been excluded by other direct detection experiments~\cite{SuperCDMS:2017mbc, SuperCDMS:2018gro, PICO:2019vsc, DarkSide-50:2022qzh, PandaX:2025rrz, XENONCollaborationP:2026ioh}, including COSINE-100~\cite{Adhikari:2018ljm, COSINE-100:2025kbw}. Because COSINE-100 uses the same target material, it can also test the DAMA/LIBRA signal without relying on such assumptions~\cite{COSINE-100:2024nfa}. A combined analysis of COSINE-100 and ANAIS-112 data excludes the DAMA/LIBRA modulation at 4.68$\sigma$ in the 1--6\,keV energy region~\cite{ANAIS-112:2025fne}.

An important aspect in comparing results among NaI(Tl)-based experiments is the quenching factor~(QF), defined as the ratio of the electron-equivalent energy to the corresponding nuclear recoil energy. Throughout this paper, keV denotes the electron-equivalent energy (keV$_\mathrm{ee}$), and nuclear recoil energies are explicitly denoted by keV$_\mathrm{nr}$. Recent measurements with NaI(Tl) crystals give QF values, approximately 0.13 for sodium and 0.05 for iodine at 10\,keV$_\mathrm{nr}$~\cite{Lee:2024unz,Collar:2013gu,Xu:2015wha,Cintas:2024pdu}, whereas DAMA/LIBRA reported significantly higher values of 0.3 for sodium and 0.09 for iodine~\cite{Bernabei:1996vj, Bernabei:2020mon}. Because of this difference, the regions of interest of the two experiments should be compared in nuclear recoil energy. The latest COSINE-100 modulation analysis considered multiple energy intervals of 1--3, 1--6, and 2--6\,keV, as well as a nuclear recoil interval of 6.7--20\,keV$_{\mathrm{nr}}$ defined for direct comparison with the DAMA/LIBRA 2--6\,keV region~\cite{COSINE-100:2024nfa}.

Extending this comparison to the lowest DAMA/LIBRA energy bins, 1--2\,keV, requires COSINE-100 to reach nuclear recoil energies below 6.7\,keV$_{\mathrm{nr}}$, and hence a lower analysis threshold. Because the two experiments calibrate their energy scales differently, the thresholds should be compared on a common scale. DAMA/LIBRA adopts a linear calibration anchored at the 59.5\,keV line of $^{241}$Am~\cite{DAMA:2008bis}, whereas COSINE-100 corrects for the nonproportional light response of NaI(Tl) at low energies~\cite{COSINE-100:2024log}; we refer to these as the linear and the corrected scales, respectively. On the linear scale, the DAMA/LIBRA threshold of 1\,keV corresponds to 3.33\,keV$_{\mathrm{nr}}$ with the DAMA/LIBRA sodium QF of 0.3~\cite{Bernabei:2020mon}, and this recoil energy in turn corresponds to 0.39\,keV with the sodium QF measured for the COSINE-100 crystals~\cite{Lee:2024unz, Cintas:2024pdu}. The 0.49\,keV threshold achieved in this work is defined on the corrected scale and corresponds to approximately 0.385\,keV on the linear scale. The COSINE-100 threshold is therefore comparable to, or slightly below, the DAMA/LIBRA threshold within the uncertainties of the QF measurements, so that the full DAMA/LIBRA energy region of interest becomes accessible to COSINE-100 modulation and spectral analyses.

In this paper, we describe the event selection procedure that enables the analysis threshold to be lowered to 0.49\,keV. The reduced threshold enables a more comprehensive comparison with DAMA/LIBRA and enhances the sensitivity to low-mass WIMPs: their recoil spectrum falls steeply with energy, so a lower threshold substantially increases the detectable fraction of their signal. In addition, since both $^{23}$Na and $^{127}$I have odd proton numbers, the improved low-energy sensitivity provides competitive constraints on spin-dependent WIMP--proton interactions at the global level, as recently demonstrated by a COSINE-100 annual modulation analysis with a few-photoelectron threshold~\cite{COSINE-100:2026flt}.

\section{The COSINE-100 Experiment}\label{sec2}

The COSINE-100 experiment~\cite{Adhikari:2017esn} is designed to search for WIMPs. The experiment was located at the Yangyang Underground Laboratory in Korea, under approximately 700\,m of rock overburden, and operated from 2016 to 2023. The detector consists of eight low-background NaI(Tl) crystals with a total mass of 106\,kg. Each crystal is cylindrical, hermetically sealed in a copper encapsulation, and optically coupled at both ends to 3-inch Hamamatsu R12669SEL photomultiplier tubes~(PMTs). The crystals have an average light yield of approximately 15\,photoelectrons/keV.

The crystal array is immersed in approximately 2{,}200\,L of liquid scintillator~(LS)~\cite{Adhikari:2020asl}. The LS shields the crystals from external radiation and tags events that also deposit energy in the LS, so that they can be removed from the DM-search sample. The LS volume is surrounded by 3\,cm of copper and 20\,cm of lead shielding to suppress external radiation backgrounds. The outermost layer of the detector is equipped with 37 plastic scintillator panels that efficiently identify cosmic-ray muons and veto associated events~\cite{COSINE-100:2017dsl}. This multi-layer shielding provides a low-background environment suitable for low-energy measurements.

Data acquisition is performed by amplifying the electrical signals from each PMT using preamplifiers and digitizing them with analog-to-digital converters operating at a sampling rate of 500\,MS/s. When the trigger condition is satisfied, a trigger board records 8\,$\mu$s waveforms, including a pre-trigger window of about 2.4\,$\mu$s. The trigger condition requires signals exceeding a height equivalent to one photoelectron in both PMTs coupled to a single crystal within a 200\,ns window~\cite{COSINE-100:2018rxe}.

Data selection begins by rejecting crystal events occurring within 30\,ms after a muon identified by the plastic scintillators. Events with coincident signals in other crystals or in the LS are classified as multiple-hit events, and only single-hit events are retained for the WIMP search. 
This selection efficiently suppresses internal radioactive backgrounds, particularly those from $^{40}$K and $^{22}$Na decays, which produce low-energy X-rays together with high-energy $\gamma$-rays. In the case of $^{40}$K, a 3.2\,keV X-ray is emitted in coincidence with a 1460\,keV $\gamma$-ray, while $^{22}$Na produces a 0.87\,keV X-ray accompanied by a 1274\,keV $\gamma$-ray. Multiple-hit events are used as a control sample to validate background models and evaluate the performance of pulse shape discrimination~(PSD). Throughout this paper, single-hit events before the event selection described in Sec.~\ref{sec3} are referred to as physics data, and those passing it as DM-search data.

The COSINE-100 experiment collected data for 6.4\,years with a data acquisition efficiency exceeding 95\,\%. Of the eight crystals, two with low light yields and one with a high PMT-induced noise rate are excluded, as in the previous analyses~\cite{COSINE-100:2024nfa}, leaving five crystals with a total effective mass of 61.3\,kg. The full dataset, hereafter referred to as the 6-year dataset, corresponds to an effective live time of 5.8\,years and an exposure of 358\,kg$\cdot$years. Detector stability was continuously monitored through real-time measurements of environmental and operational parameters, including temperature, humidity, radon concentration, PMT high voltage, and current~\cite{COSINE-100:2021mlj}. 
Long-term gain correction was performed by tracking the 46.5\,keV internal peak from $^{210}$Pb decays, which served as the primary reference for monitoring and correcting time-dependent PMT gain shifts~\cite{COSINE-100:2021zqh}. 
Energy calibration accounting for the nonproportional light response of NaI(Tl) at low energies was performed using multiple internal peaks, including the 3.2\,keV peak from $^{40}$K and the 0.87\,keV peak from $^{22}$Na~\cite{COSINE-100:2024log}.

The recorded waveforms are analyzed using PSD techniques to distinguish scintillation events from PMT-induced noise events, hereafter referred to as noise. Various PSD parameters are constructed, including the mean time, pulse slope, head-to-tail charge ratios, and likelihood-based discrimination variables. A boosted decision tree~(BDT) trained with these input PSD parameters enabled the analysis threshold for the initial 1.7-year dataset to be lowered to 1\,keV~\cite{COSINE-100:2020wrv}. Subsequently, with improved likelihood-based parameters and a multi-layer perceptron~(MLP) network, the analysis threshold for the 3-year dataset was reduced to 0.7\,keV~\cite{COSINE-100:2024wji}. In this work, we lower the threshold further for the 6-year dataset.

\section{Event selection with a multi-layer perceptron}\label{sec3}

For the event selection of the 6-year dataset, we use an MLP implemented within the Toolkit for Multivariate Data Analysis~(TMVA)~\cite{TMVA:2007ngy}, similar to that used in the previous 3-year analysis~\cite{COSINE-100:2024wji}. Two major improvements are introduced to enhance the PSD performance, particularly in the low-energy region below 1\,keV: the use of simulated signal waveforms for training, and a separate training for each group of events classified by the number of clusters~(NC) in the waveform (NC group).

The first improvement is the use of simulated signal waveforms. The performance of an MLP depends on the size of its training samples~\cite{Goodfellow2016}. In the previous analysis~\cite{COSINE-100:2024wji}, scintillation events selected from $^{22}$Na-calibration data were used as the signal sample, which was much smaller than the noise sample. In this work, signal events were generated with a waveform simulation~\cite{Choi:2024ziz}, so that the signal sample could be as large as needed. For the noise sample, approximately 20\,\% of the 6-year physics data were used after pre-selection cuts; with the doubled live time, this sample is also larger than before. The noise sample also contains scintillation events; 1.0 to 1.7\,\% of its events with fewer than 15 NC pass the event selection. A separate MLP was trained for each crystal and NC group, using $(2.6$ to $5.4)\times10^{5}$ simulated signal events and $(2.6$ to $7.6)\times10^{5}$ noise-sample events per group. The same pre-selection cuts were applied to both samples so that they cover the same region of the input-variable space.

\begin{figure}[htbp]
\centering
\includegraphics[width=0.8\linewidth]{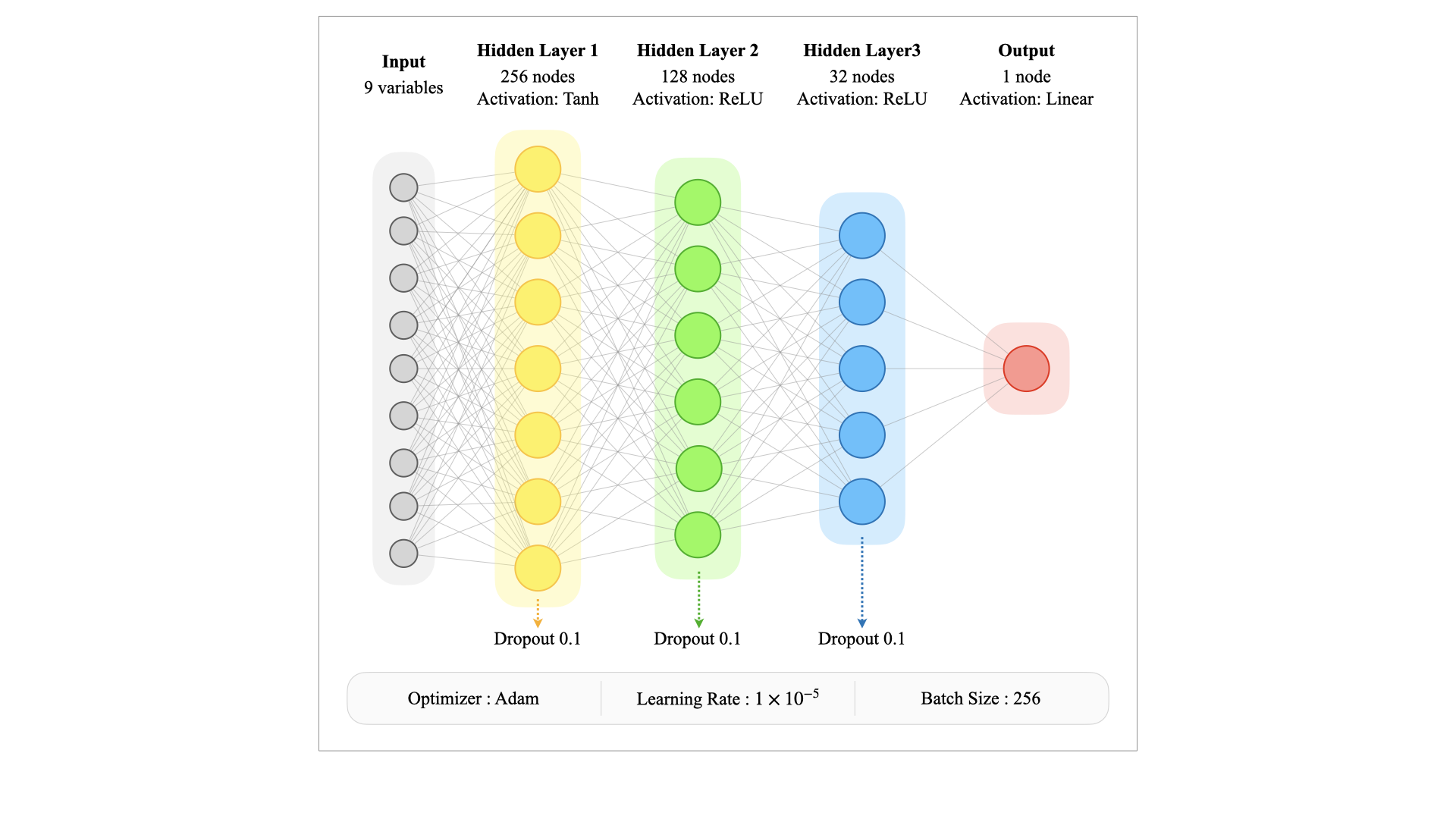}
\caption{
Schematic diagram of the MLP architecture used for event classification.
}
\label{fig:MLPschematic}
\end{figure}

Figure~\ref{fig:MLPschematic} shows a schematic diagram of the MLP architecture. The MLP has three hidden layers with 256, 128, and 32 nodes, respectively, followed by a single output node with a linear activation function. The first hidden layer uses the hyperbolic tangent activation function, while the remaining hidden layers use rectified linear unit~(ReLU) activation functions~\cite{Nair:2010rel, Glorot2011}. 
The network was trained by minimizing the cross-entropy loss, for which the linear output is mapped to a score between 0 and 1 by a sigmoid function. 
The model was optimized using the Adam algorithm~\cite{Kingma:2014vow} with $\beta_1=0.9$, $\beta_2=0.999$, and $\epsilon=10^{-7}$, a learning rate of $1\times10^{-5}$, and a batch size of 256, for up to 1000 epochs. To prevent overfitting, $L_2$ regularization with a weight decay of $1\times10^{-4}$ and a dropout rate of 0.1 were applied to all hidden layers, and early stopping was applied when the performance did not improve for 30 consecutive epochs. Overfitting was checked by comparing the training and validation errors for each crystal and NC group; the validation error exceeds the training error by at most 0.012, indicating no significant overfitting.

The MLP takes nine input variables that characterize the pulse shape of each event. These variables are defined in detail in the previous analysis~\cite{COSINE-100:2024wji}, and the same set is adopted here without modification. Following Refs.~\cite{COSINE-100:2020wrv, COSINE-100:2024wji}, noise events are classified into two types: Type-1 noise, characterized by large spikes and short mean times, and Type-2 noise, characterized by small spikes and long mean times. Type-2 noise is more difficult to distinguish from scintillation events because its long-tailed structure resembles that of scintillation pulses. A brief description of each variable is given below.
Two charge-ratio parameters are used to capture the temporal structure of the scintillation pulse. The first, $Q_\mathrm{tail}/Q_\mathrm{total}$, is the ratio of the charge integrated in the tail region to the total charge, which is sensitive to the slower decay component of scintillation events relative to noise. The second, $Q_\mathrm{head}/Q_\mathrm{total}$, is the ratio of the charge in the prompt region, which is sensitive to fast-rising noise (Type-1).
Two mean-time parameters are included. The PMT-based mean-time is computed separately for each PMT waveform and primarily targets Type-1 noise. The summed mean-time is computed from the two-PMT summed waveform per crystal, providing complementary discrimination power.
Five likelihood-based parameters are used to exploit the full waveform shape. The likelihood parameter is constructed using a Type-1 noise template and targets Type-1 noise events. The unbinned likelihood parameter and the PMT-based likelihood parameter are both constructed using Type-2 noise templates, targeting Type-2 noise. The two Fourier likelihood parameters are computed from Fourier-transformed waveforms, targeting Type-1 and Type-2 noise, respectively.

\begin{figure}[htbp]
\centering
\begin{tabular}{cc}
\includegraphics[width=0.45\linewidth]{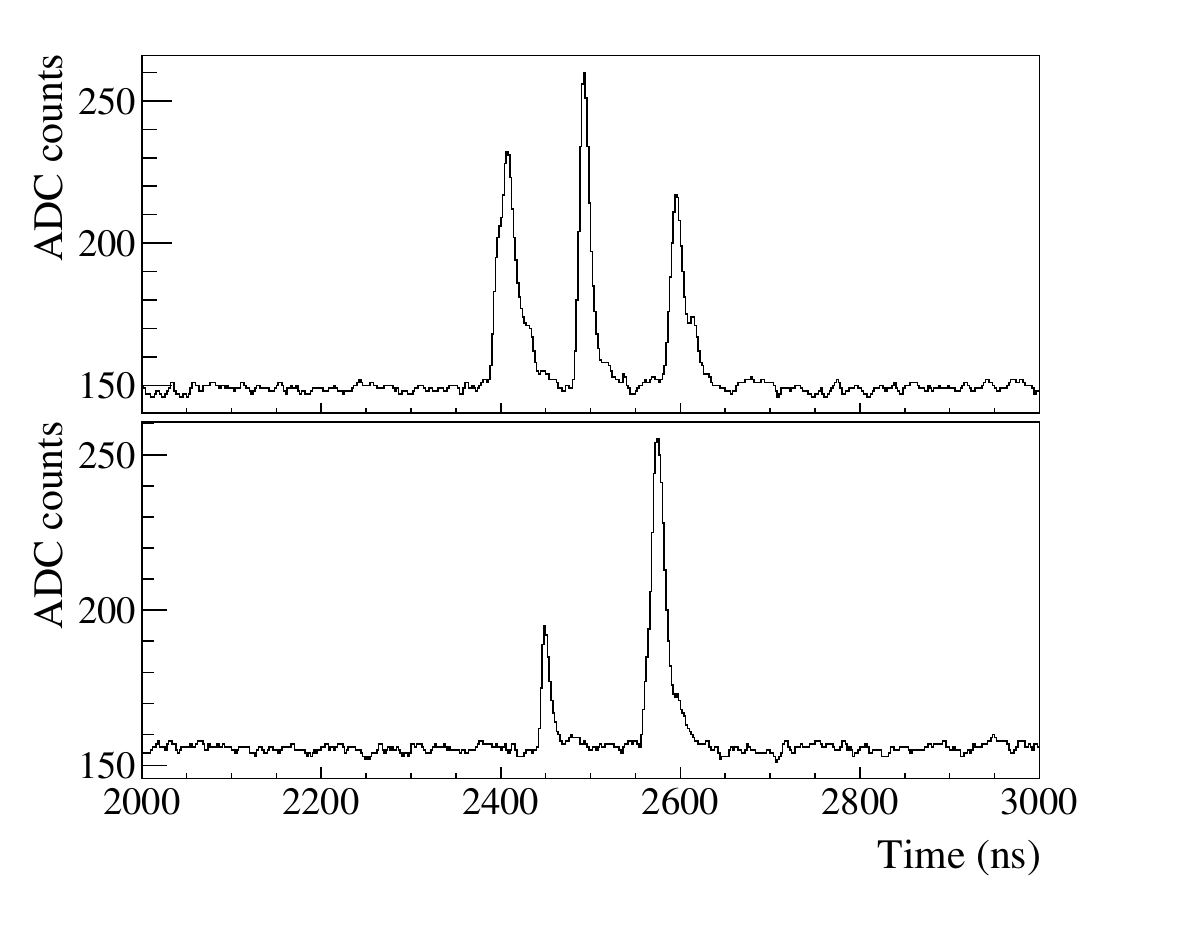} &
\includegraphics[width=0.45\linewidth]{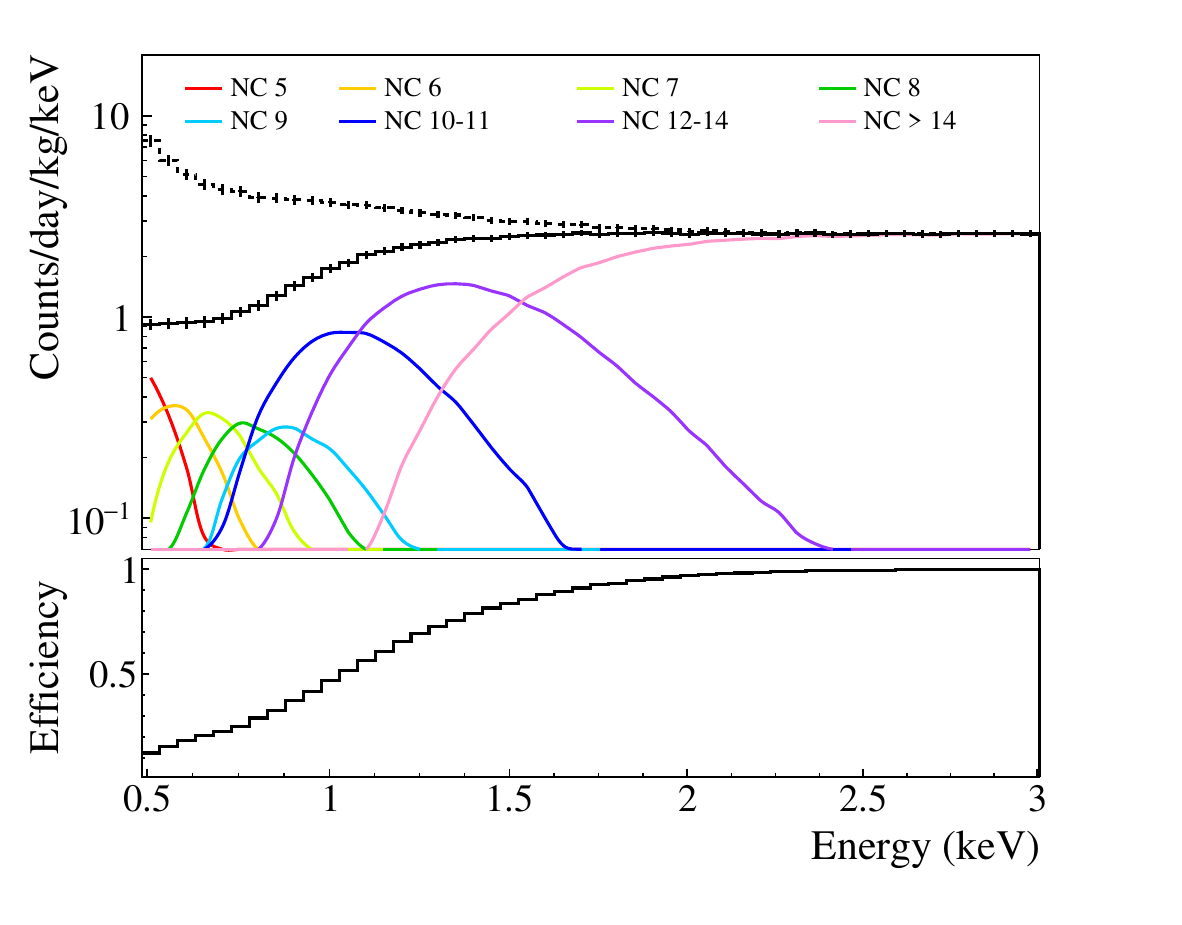} 
\end{tabular}
\caption{
(Left) Waveforms of a scintillation event with NC\,=\,5 recorded by PMT1 (top, three NC) and PMT2 (bottom, two NC). (Right) Energy spectra of the DM-search data, i.e., single-hit events passing the event selection, averaged over the five crystals. The colored lines show the contributions of the eight NC groups, and the black solid line shows the total spectrum. The black dashed line shows the spectrum corrected for the selection efficiency, which is shown in the bottom panel and is the efficiency applied in the WIMP search.
}
\label{fig:NCenergy}
\end{figure}

The second improvement is a separate training for each NC group. To optimize the performance in the low-energy region, events were categorized according to NC, defined as the number of isolated peaks in the waveform that exceed a pedestal-based threshold, identified with the clustering algorithm of Ref.~\cite{Kims:2005dol} as in Ref.~\cite{COSINE-100:2026flt}. 
An example waveform of a scintillation event with NC = 5, at an energy of approximately 0.5\,keV, is shown in the left panel of Fig.~\ref{fig:NCenergy}, illustrating how the total NC is obtained by summing the cluster counts from the two PMTs. 
Events with four or fewer NC were removed by the pre-selection because of their very low energies. The remaining events were divided into eight NC groups, whose mean energies are listed in Table~\ref{table:NCenergy} and whose energy spectra are shown in Fig.~\ref{fig:NCenergy}. The seven groups with NC below 15 mostly cover energies below about 2\,keV, while all events with 15 or more NC, up to 6\,keV, are combined into a single group. The event selection is therefore applied to the energy range from the threshold up to 6\,keV, while events above 6\,keV are selected with the event selection developed for the 1\,keV threshold analysis~\cite{COSINE-100:2020wrv}.
\begin{table}[h]
\caption{Mean energy of event groups defined by NC.}\label{table:NCenergy}%
\begin{tabular}{@{}lllllllll@{}}
\toprule
Group & 1 & 2 & 3 & 4 & 5 & 6 & 7 &8 \\
\midrule
NC & 5 & 6 & 7 & 8 & 9 & 10--11 & 12--14 & $\geq$15\\
$\langle E \rangle$ (keV) & 0.53 & 0.64 & 0.75 & 0.87 & 1.01 & 1.23 & 1.59 & 3.66\\
\botrule
\end{tabular}
\end{table}

\begin{figure}[htbp]
\centering
\includegraphics[width=0.95\linewidth]{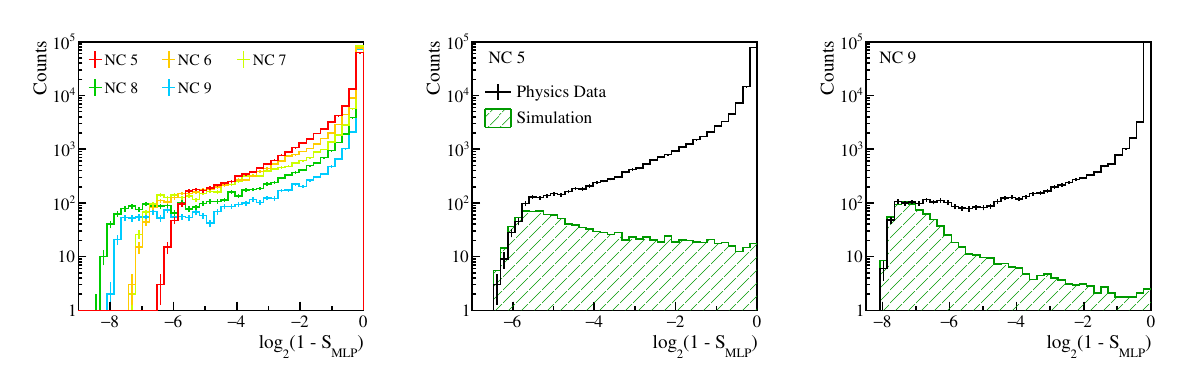}
\caption{
Distributions of $\log_2(1-S_{\mathrm{MLP}})$ for crystal 6, shown as an example, restricted to events below 1\,keV, where the discrimination is most difficult. (Left) Physics data for NC 5 to 9. (Center and right) Physics data (black) compared with the scaled waveform simulation (green hatched) for NC 5 and NC 9.
}
\label{fig:mlp_score}
\end{figure}

The resulting MLP score~$(S_{\text{MLP}})$ distributions are shown in Fig.~\ref{fig:mlp_score}, where the horizontal axis is expressed as $\log_2(1-S_{\text{MLP}})$ to better resolve the high-score (signal-dominated) region. 
The $S_{\text{MLP}}$ distribution of the waveform simulation reproduces the physics data closely in the region of low $\log_2(1-S_{\text{MLP}})$ values, indicating that the data in this region are consistent with scintillation events. 
As $\log_2(1-S_{\text{MLP}})$ increases, the data distribution begins to exceed the scaled simulation, and this excess is interpreted as residual noise contamination. The event selection criterion for each NC group was determined by requiring this noise contamination level to remain below 1\,\%. The selection efficiency is defined as the fraction of simulated scintillation events that pass the selection criterion. The analysis threshold is set to the lowest energy at which the selection efficiency remains above 10\,\%, which is 0.49\,keV.

\section{Validation}\label{sec4}

We validate the event selection in three steps. First, we check that the simulation reproduces the input variables of scintillation events in data. Second, we test the stability of the selection over the six years of data taking. Third, we compare the simulated selection efficiency with efficiencies measured with two independent data samples.

\begin{figure}[htbp]
\centering
\includegraphics[width=0.95\linewidth]{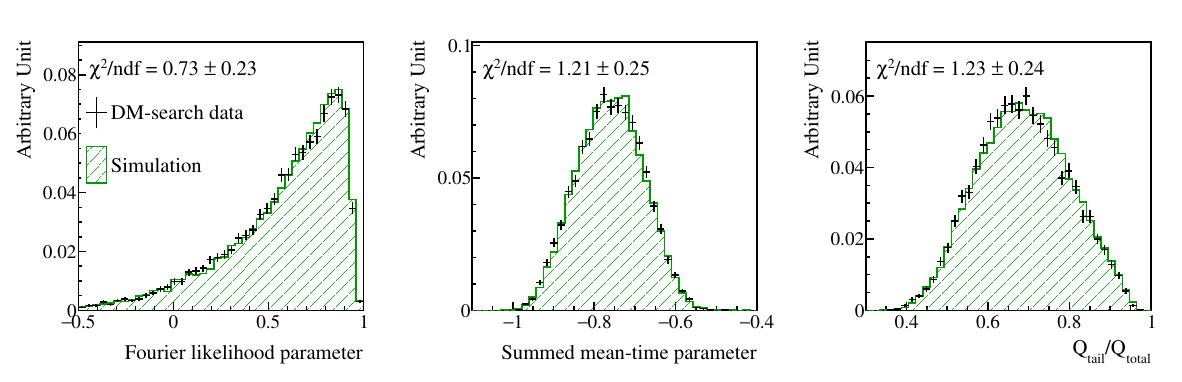}
\caption{
Probability density functions of representative input variables for crystal 2: the Fourier likelihood parameter for Type-1 noise (left), the summed mean-time parameter (center), and $Q_{\mathrm{tail}}/Q_{\mathrm{total}}$ (right). DM-search data (black markers) are compared with simulation (green shaded histograms), each normalized to unit area.
}
\label{fig:validation_variables}
\end{figure}

To validate the modeling of scintillation events used for training, we compare the probability density functions~(PDFs) of key input variables between DM-search data and simulation.
The variables considered are the Fourier likelihood parameter for Type-1 noise, the summed mean-time parameter, and the charge integration ratio $Q_{\mathrm{tail}}/Q_{\mathrm{total}}$.
For each variable, the agreement between data and simulation is quantified using a $\chi^{2}$ test.
As shown in Fig.~\ref{fig:validation_variables}, good agreement is observed across all variables, with $\chi^{2}/\mathrm{ndf}$ values close to unity.
This demonstrates that the simulation reliably reproduces the characteristics of scintillation events in the DM-search data.

\begin{figure}[htbp]
\centering
\includegraphics[width=0.95\textwidth]{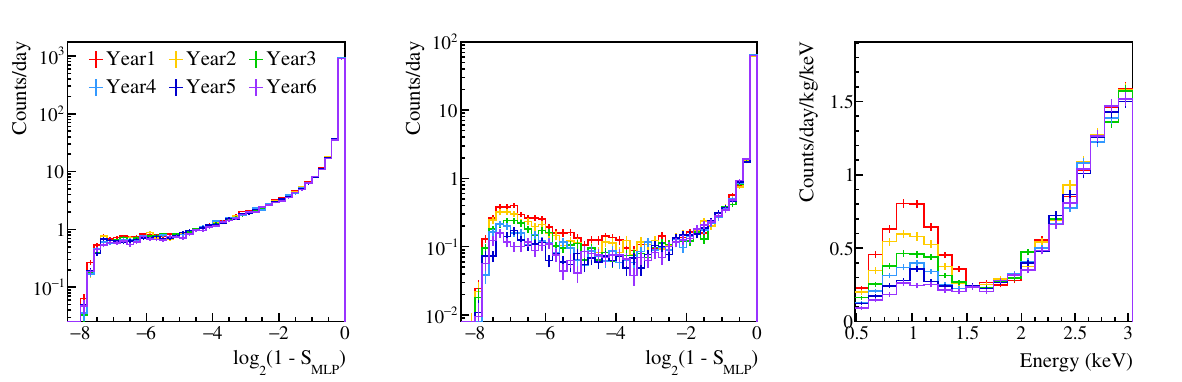} 
\caption{
Time dependence for crystal 4, shown as an example, over six data-taking periods of approximately one year each. (Left and center) Distributions of $\log_{2}(1-S_{\text{MLP}})$ for single-hit and multiple-hit physics data with NC\,=\,8. (Right) Low-energy spectra of all multiple-hit events passing the event selection, without restriction on NC. The decrease in the signal-like region of the multiple-hit distribution (center) corresponds to the decrease of the internal $^{22}$Na peak near 0.87\,keV (right).
}
\label{fig:time_stability}
\end{figure}

We next evaluate the long-term stability of the event selection by dividing the full 6-year dataset into six periods of approximately 1\,year each. The distributions of $\log_2(1-S_{\mathrm{MLP}})$ for the six periods are shown in Fig.~\ref{fig:time_stability} (left). The six distributions are consistent within statistical uncertainties, indicating no significant time dependence in the selection performance.

In addition, the low-energy multiple-hit energy spectra for the six periods are presented in Fig.~\ref{fig:time_stability}. A peak around 0.87\,keV is clearly visible and gradually decreases with time, as expected from the decay of internal $^{22}$Na. To verify this, we select internal $^{22}$Na events from the multiple-hit data passing the event selection by requiring an energy deposition below 2\,keV in one crystal and a coincident energy deposition of 1150--1350\,keV, corresponding to the 1274\,keV $\gamma$-ray from $^{22}$Na decay, in another crystal. The yearly rates of these events are well described by an exponential decay, and the fitted half-lives are consistent with the $^{22}$Na half-life of 2.6\,years within uncertainties for all five crystals. This time-dependent feature demonstrates that the event selection reliably identifies scintillation events even in the sub-keV energy region.

\begin{figure}[htbp]
\centering
\includegraphics[width=0.6\linewidth]{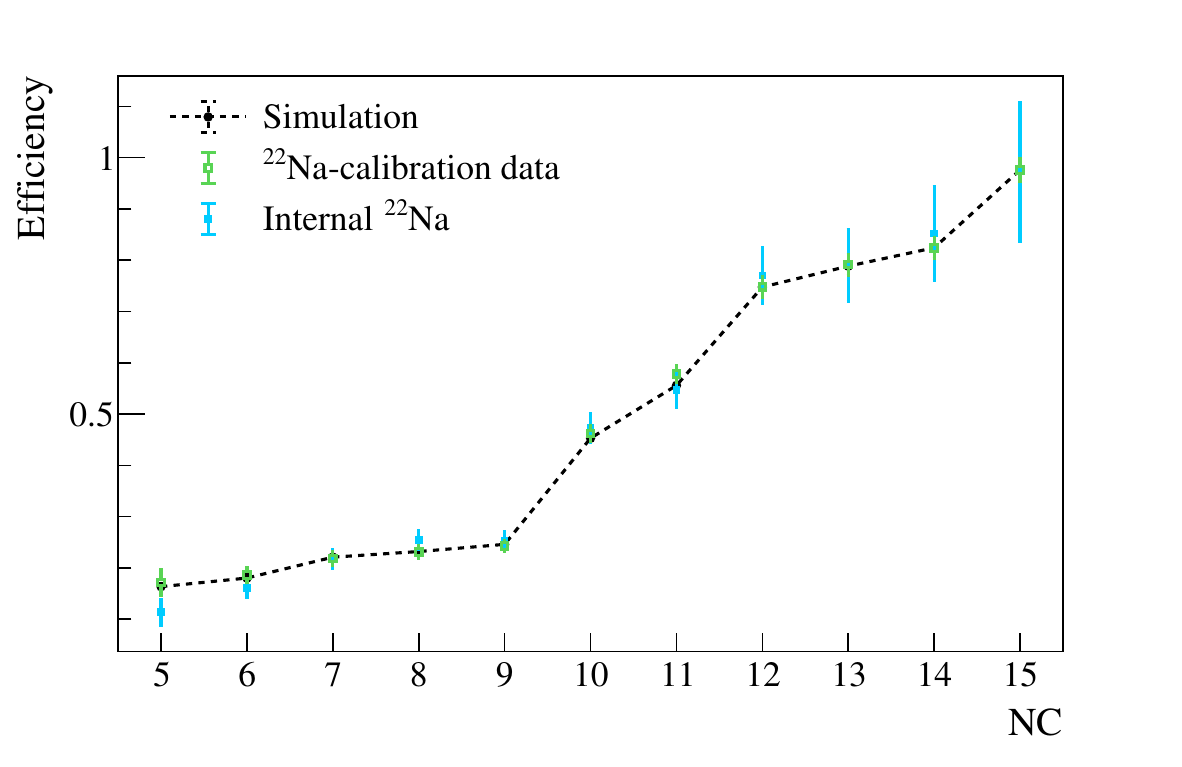}
\caption{
Event selection efficiency as a function of NC, averaged over the five crystals, derived from simulation (black), $^{22}$Na-calibration data (green), and internal $^{22}$Na events (blue). The internal $^{22}$Na sample consists of multiple-hit events below 2\,keV with a coincident energy deposition of 1150--1350\,keV in another crystal, taken before the event selection is applied. Error bars indicate statistical uncertainties.
}
\label{fig:efficiency_validation}
\end{figure}

To validate the selection efficiency derived from simulation, we compare it with efficiencies independently estimated using two data-driven approaches, as shown in Fig.~\ref{fig:efficiency_validation}. 
The first method uses $^{22}$Na-calibration data, from which a scintillation-enriched sample is selected using pre-selection cuts; this dataset was also employed as the signal training sample for the MLP in the previous analysis~\cite{COSINE-100:2024wji}. 
The second method uses internal $^{22}$Na events selected with the same coincidence condition but before the event selection is applied; the efficiency is given by the fraction of these events that pass the event selection. This coincidence requirement ensures that the selected low-energy events are predominantly scintillation events, providing a clean, scintillation-enriched validation sample.
These internal events provide an in-situ validation sample that reflects interactions occurring uniformly throughout the crystal volume, similar to WIMP-induced nuclear recoils.

The efficiencies obtained from both the calibration data and internal $^{22}$Na events are consistent with the simulation-based efficiency within statistical uncertainties, demonstrating the robustness of the efficiency estimation. The simulation-based efficiency is used as the nominal value. Its statistical uncertainty, arising from the size of the simulated sample, and its differences from the efficiencies obtained with the two validation samples are assigned as systematic uncertainties of the selection efficiency, which are propagated to the sensitivity estimate in Sec.~\ref{sec:sensitivity}.

\section{Sensitivity}\label{sec:sensitivity}

\begin{figure}[htbp]
\centering
\includegraphics[width=0.6\textwidth]{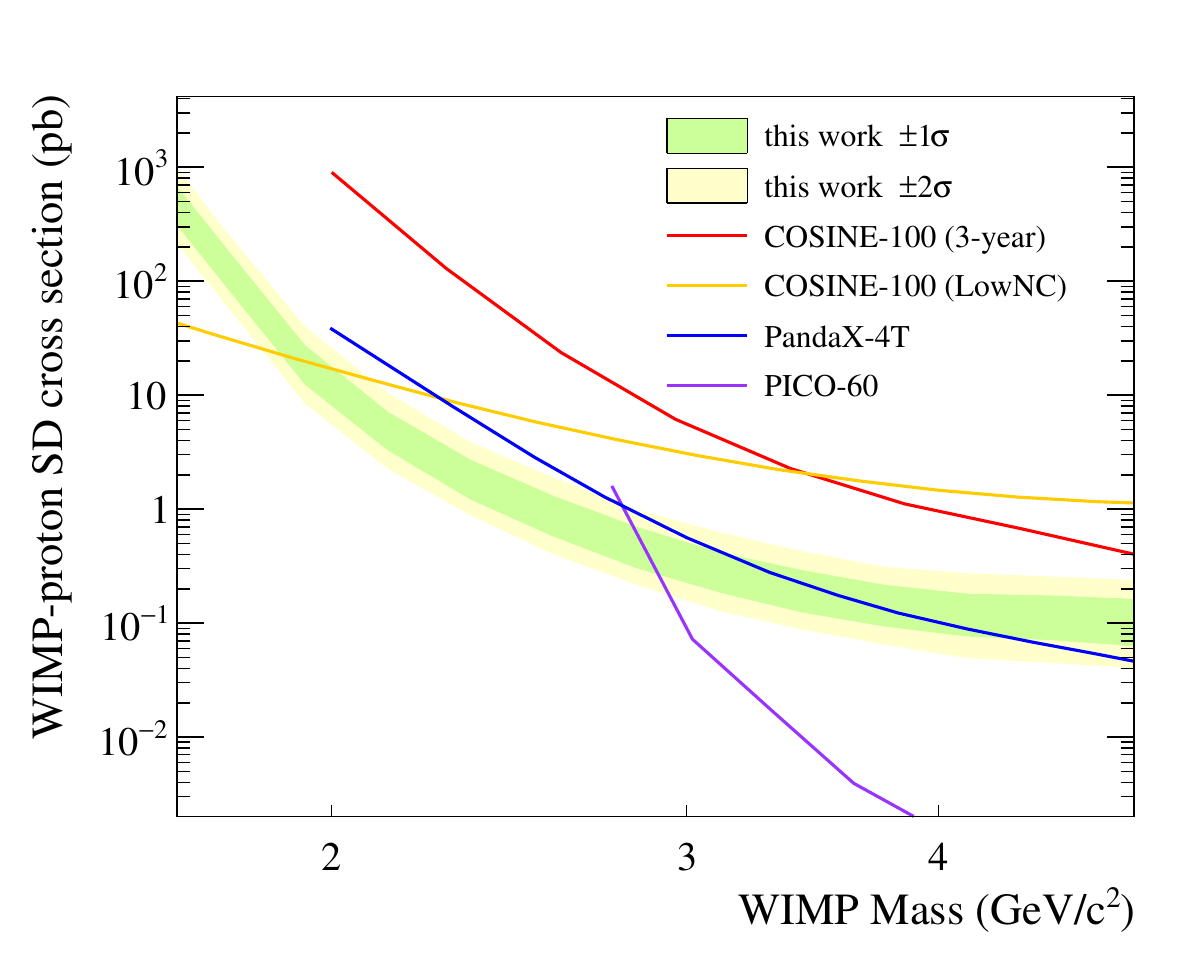}
\caption{
Expected 90\,\% credible upper limits on the SD WIMP--proton cross section for the full 6-year COSINE-100 dataset with a 0.49\,keV analysis threshold, derived from 1,000 pseudo-experiments fitted in the 0.49--15\,keV energy range. The green and yellow bands represent the $\pm1\sigma$ and $\pm2\sigma$ ranges of the expected limits, respectively. The limits from the previous COSINE-100 analyses of the 3-year dataset~\cite{COSINE-100:2025kbw} and of low-NC events~\cite{COSINE-100:2026flt}, PandaX-4T~\cite{PandaX:2025rrz}, and PICO-60~\cite{PICO:2019vsc} are shown for comparison.
}
\label{fig:sensitivity}
\end{figure}

The lower analysis threshold, together with the improved selection efficiency and noise rejection at sub-keV energies, directly benefits searches for low-mass WIMPs. Because the nuclear recoil spectrum of a light WIMP falls steeply with energy and is further suppressed by quenching, a large fraction of its signal lies below 1\,keV, where the selection efficiency of the previous analysis with a 0.7\,keV threshold was limited~\cite{COSINE-100:2024wji}. We quantify this gain for spin-dependent~(SD) WIMP--proton interactions, to which NaI(Tl) is intrinsically sensitive because both $^{23}$Na and $^{127}$I have odd proton numbers, and for which the light $^{23}$Na nucleus provides kinematic access to low-mass WIMPs. Figure~\ref{fig:sensitivity} shows the expected sensitivity of the full 6-year dataset for WIMP masses of 1.7--5\,GeV/$c^{2}$.

The sensitivity is evaluated with pseudo-experiments generated in the energy range from the 0.49\,keV threshold up to 15\,keV, which is the same range used in the fit of the DM-search data; the present event selection is applied below 6\,keV and that of Ref.~\cite{COSINE-100:2020wrv} above 6\,keV. The background model of the 3-year analysis~\cite{COSINE-100:2024ola} is adopted and rescaled to the exposure and selection efficiency of the present dataset. For each pseudo-experiment, the activity of each background component is sampled from a Gaussian distribution whose width corresponds to its uncertainty in the model, and the systematic uncertainties, including those of the selection efficiency described in Sec.~\ref{sec4}, energy calibration, energy resolution, and background shapes, are sampled in the same manner. The resulting expected spectrum is then fluctuated bin-by-bin according to Poisson statistics to account for statistical uncertainties. A total of 1,000 pseudo datasets are generated in this way.

Each pseudo dataset is fitted with a Bayesian approach using a model consisting of the background components, the systematic uncertainties treated as nuisance parameters, and a WIMP signal. The WIMP-induced nuclear recoil spectra are calculated with the \texttt{dmdd} package~\cite{Gluscevic:2015ascl, Gluscevic:2015sqa}, which incorporates the nuclear response functions of Refs.~\cite{Fitzpatrick:2012ix, Anand:2013yka, Gresham:2014vja}. The WIMP signal is modeled assuming the standard halo model with a local dark matter density of 0.3\,GeV/cm$^{3}$, $v_{0}=238$\,km/s, $v_{\oplus}=250.55$\,km/s, and $v_{\mathrm{esc}}=544$\,km/s, and the recoil energies are converted to electron-equivalent energies using the QF measured for the COSINE-100 crystals~\cite{Lee:2024unz}. The posterior distribution of the WIMP--proton cross section is obtained by marginalizing over all other parameters, from which the 90\,\% credible upper limit is determined for each pseudo dataset. The distribution of these upper limits defines the expected sensitivity, with the $\pm1\sigma$ and $\pm2\sigma$ bands given by the central 68\,\% and 95\,\% intervals, respectively.

The expected sensitivity is compared with the previous COSINE-100 results from the 3-year dataset~\cite{COSINE-100:2025kbw} and from the low-NC analysis~\cite{COSINE-100:2026flt}, as well as with the limits from PandaX-4T~\cite{PandaX:2025rrz} and PICO-60~\cite{PICO:2019vsc}. The low-NC analysis searched for an annual modulation signal using events with 2--4 NC from a part of the 6-year dataset, whereas the present event selection starts from NC\,=\,5 (Table~\ref{table:NCenergy}). In the mass range of approximately 2.0--2.8\,GeV/$c^{2}$, the median expected limit lies below all of these existing limits, by up to a factor of about 3.5 near 2.3\,GeV/$c^{2}$, and reaches 1.1\,pb ($1.1 \times 10^{-36}$\,cm$^{2}$) at 2.5\,GeV/$c^{2}$. In this mass range, the expected limit is also a factor of 16 to 70 below the 3-year COSINE-100 result, reflecting the lower threshold, the higher selection efficiency near threshold, and the increased exposure. Above 2.8\,GeV/$c^{2}$, the PICO-60 limit, obtained with a fluorine target, is more stringent. The full 6-year COSINE-100 dataset is thus expected to explore previously unexplored parameter space for SD WIMP--proton interactions in the 2.0--2.8\,GeV/$c^{2}$ mass range.

\section{Conclusion}

We have presented an improved event selection strategy for the COSINE-100 experiment based on a machine learning approach, enabling the analysis threshold to be reduced to 0.49\,keV using the full 6-year dataset. This threshold is comparable to that of DAMA/LIBRA once the differences in energy calibration and QF are taken into account, allowing COSINE-100 analyses to cover the full DAMA/LIBRA energy region of interest.

The improved performance is achieved through two key developments: the introduction of waveform-simulated signal samples and a separate training for each NC group. These enhancements significantly improve the discrimination between scintillation events and noise, particularly in the low-energy region. The resulting event selection achieves a signal efficiency exceeding 10\,\% at 0.49\,keV while maintaining noise contamination below 1\,\%.

The method is validated in three ways: the simulation reproduces the input variables of DM-search data, the selection performance is stable over the six years of data taking, and the simulated efficiency agrees with efficiencies measured with $^{22}$Na-calibration data and internal $^{22}$Na events.

Based on this improved event selection, the expected sensitivity to WIMP interactions is evaluated using pseudo-experiments incorporating statistical fluctuations and systematic uncertainties. The reduced threshold significantly enhances the sensitivity in the low-mass region. For spin-dependent WIMP--proton interactions, COSINE-100 is expected to probe previously unexplored parameter space in the 2.0--2.8\,GeV/$c^{2}$ mass range, with a median expected limit of $1.1 \times 10^{-36}$\,cm$^{2}$ at 2.5\,GeV/$c^{2}$.

These results show that a NaI(Tl) detector with a sub-keV threshold can provide competitive constraints on low-mass WIMPs, in particular for spin-dependent WIMP--proton interactions, and can extend NaI(Tl)-based tests of DAMA/LIBRA to its full energy region of interest.

\bmhead{Acknowledgements}
We thank the Korea Hydro and Nuclear Power~(KHNP) Company for providing underground laboratory space at Yangyang and the IBS Research Solution Center~(RSC) for providing high performance computing resources.
This work is supported by: 
the Institute for Basic Science~(IBS) under project code IBS-R016-A1, NRF-2021R1A2C3010989, NRF-2021R1A2C1013761, RS-2024-00356960, RS-2025-25442707 and RS-2025-16064659, Republic of Korea; 
NSF Grants No. PHY-1913742, United States; 
STFC Grant ST/N000277/1 and ST/K001337/1, United Kingdom; Grant No. 2021/06743-1, 2022/12002-7, 2022/13293-5 and 2025/01639-2 FAPESP, CAPES Finance Code 001, CNPq 304658/2023-5, Brazil; 
UM Internal Grant non-APBN 2025, Indonesia.

\bibliography{sn-bibliography}

\end{document}